\documentclass[11pt]{article}

\usepackage[]{acl}

\usepackage{times}
\usepackage{latexsym}
\usepackage{multicol}
\usepackage{multirow}

\usepackage[T1]{fontenc}

\usepackage[utf8]{inputenc}

\usepackage{microtype}
\usepackage[most]{tcolorbox}
\usepackage{listings}
\lstdefinestyle{json}{
  basicstyle=\ttfamily\small,
  columns=fullflexible,
  breaklines=true,
  breakatwhitespace=true,
  frame=none,
  showstringspaces=false
}

\usepackage{inconsolata}
\newcommand{\checkbox}{\fbox{\rule{0pt}{0.5ex}\rule{0.5ex}{0pt}}}

\usepackage{graphicx}

\title{Commitment Checklist: Auditing Author Commitments in Peer Review}

\author{
  \textbf{Chung-Chi Chen\textsuperscript{1}},
  \textbf{Iryna Gurevych\textsuperscript{2}}
  \\
  \\
  \textsuperscript{1}AIST, Japan \\
  \textsuperscript{2}Ubiquitous Knowledge Processing Lab (UKP Lab), Department of Computer Science, \\ TU Darmstadt and National Research Center for Applied Cybersecurity ATHENE, Germany \\
  \\
  \\
  \small{
    \href{mailto:c.c.chen@acm.org}{c.c.chen@acm.org},
    \href{mailto:iryna.gurevych@tu-darmstadt.de}{iryna.gurevych@tu-darmstadt.de}
  }
}

\begin{document}
\maketitle
\begin{abstract}
Peer review author responses often include commitments to add experiments, release code, or clarify content in the final paper. Yet, there is currently no systematic mechanism to ensure authors fulfill these promises. In this position paper, we present a large-scale \textit{audit of author commitments} using large language models (LLMs) to compare rebuttals against camera-ready versions. Analyzing the commitments from ICLR-2025 and EMNLP-2024, we find that while a majority of promised changes are implemented, a significant share (about 25\%) are not, with ``missing experiments'' and other high-impact items among the most frequently unfulfilled. We demonstrate that LLM-based tools can feasibly detect the promises. Finally, we propose the idea of Author Commitment Checklist, which would alert authors and organizers to unaddressed promises, increasing accountability and strengthening the integrity of the peer review process. We discuss the benefits of this practice and advocate for its adoption in future conferences.
\end{abstract}

\section{Introduction}
During the rebuttal phase of peer review, authors commonly pledge to make specific improvements in their final manuscript in response to reviewer feedback. For example, an author might write: \textit{“We will run an additional experiment on dataset X and include the results in the camera-ready version.”} Such \textbf{author commitments} are intended to address reviewers’ concerns and can influence the decision process. However, once a paper is accepted, there is typically no formal mechanism to enforce these promised changes in the camera-ready submission. Program committees and reviewers rarely revisit papers post-acceptance, and currently "there is no technical requirement for the authors to fully address all reviewer suggestions after the manuscript has been accepted."~\cite{Afzal2024PeerReviewSurvey}. This lack of accountability can undermine trust in the peer review process, as reviewers might wonder if their feedback was genuinely incorporated or merely acknowledged to gain acceptance.

The community has begun to recognize this gap. In response, some venues have experimented with conditional acceptance and shepherding, in which authors are required to implement specific changes under supervision. Historically, conference policies often discouraged authors from making brand-new result promises in rebuttals, precisely because such claims could not be verified at the time of decision. Recently, however, the community has become more open to, and even encouraging of, richer discussion and additional results during the rebuttal phase. In practice, many authors now make extensive commitments in an effort to persuade reviewers. As we will show, modern conference rebuttals contain numerous promises, ranging from adding experiments and proofs to releasing code and improving the quality of writing.

Manually tracking whether tens of thousands of such commitments are fulfilled is infeasible. This is where recent advances in NLP and LLMs offer an opportunity. Large language models are increasingly being used for document comparison and consistency checking tasks. For instance, LLMs can verify if an abstractive summary aligns with a source document or if facts remain consistent after edits \cite{luo2023chatgpt}. In the context of peer review, NLP techniques have been proposed to align review comments with manuscript revisions \cite{Kuznetsov2022Revise}, and to detect new content in camera-ready versions \cite{Jiang2022ArxivDiff}. Building on these developments, we explore whether LLMs can effectively identify when an author’s promised change does \textit{not} appear in the final paper.

In this paper, we present the first empirical study of author commitments and their fulfillment. We compile a novel dataset of author response commitments from two major 2024--2025 NLP/ML conferences, ICLR-2025 and EMNLP-2024~\cite{dycke-etal-2023-nlpeer}. Using an automated pipeline, we extract and categorize these commitments, then employ well-performing LLMs to verify each against the camera-ready version of the paper. Our analysis provides quantitative insights into:
\begin{itemize}\setlength\itemsep{0em}
\item \textbf{Prevalence of commitments:} How many promises do authors make on average, and how does this differ by conference?
\item \textbf{Fulfillment rates:} What fraction of commitments are actually fulfilled in the final papers?
\item \textbf{Nature of broken promises:} Which types of commitments are most often unfulfilled? Are they primarily major additions (experiments, theory) or minor fixes?
\item \textbf{Feasibility of automated auditing:} How accurately can LLMs detect fulfilled vs.\ unfulfilled commitments, compared to human judgment?
\end{itemize}

Our results confirm that a non-trivial portion of author promises are not kept, especially those involving substantial new work. For example, we find that adding missing experiments or comparisons is the most common commitment and also among the most frequently unmet. Encouragingly, we also find that LLM-based checks can identify a majority of these cases automatically, hinting at a practical tool for oversight.

Finally, we propose a concrete solution to improve accountability: an \textbf{Author Commitment Checklist (ACC)}. ACC can be automatically generated with LLMs wherein commitments extracted from the rebuttal are presented as a checklist to authors and verified at camera-ready submission time. We outline how ACC would integrate into existing workflows and discuss its potential benefits, such as discouraging insincere promises and ensuring published papers deliver on the improvements reviewers expect. We argue that adopting ACC in future conferences can bolster the credibility of peer review without unduly burdening authors or organizers.

\section{Related Work}
Recent years have seen a push towards greater transparency in peer review, including open reviews and public discussion forums (e.g., OpenReview for ICLR) and the publication of review process reports \cite{Shah2018NeurIPSReport}. However, these efforts largely focus on the decision process and reviewer behavior, not on authors' post-decision actions. One related development has been the introduction of reproducibility and ethics checklists in the submission process \cite{Pineau2021ReproducibilityChecklist}. For instance, NeurIPS and some ACL conferences now require authors to fill out a checklist about code and data sharing. These checklists aim to encourage certain commitments (like releasing code or detailing evaluation settings) at submission time. Prior work has shown that, despite the introduction of reproducibility and reporting checklists, compliance with recommended practices remains inconsistent \cite{Dodge2019ChecklistNLP}. These checklists encourage commitments at submission time, but do not explicitly enforce that promised actions (e.g., providing code) are carried out after acceptance.
Another relevant thread is community-led post-publication verification~\cite{Pineau2021ReproducibilityChecklist}. If authors fail to deliver on promises (such as releasing code or datasets), it may be discovered and noted by readers or replicators. However, this happens after publication and is unsystematic.
Our proposed ACC can be seen as part of the broader movement toward transparent and responsible research communication. By formally recording and checking commitments before publication, ACC would increase the visibility of whether authors follow through on what was stated during review, adding a missing piece to review transparency.

\section{Methodology}
Our study involves three main components: (1) data collection from two conferences, (2) a commitment extraction and categorization pipeline, and (3) an LLM-based verification system with human validation. We describe each in detail below.

\subsection{Data Sources}
We focus on two prominent venues: ICLR 2025 (International Conference on Learning Representations) and EMNLP 2024 (Conference on Empirical Methods in Natural Language Processing). These venues were selected because both included an author response (rebuttal) phase, allowing authors to address reviewer comments, yet they differ substantially in their review transparency and interaction format. ICLR uses the OpenReview platform, where reviews and the subsequent discussion between authors and reviewers are publicly accessible. In contrast, the author responses in EMNLP 2024 and the review discussions are not publicly released after the conference.

For ICLR 2025, we accessed all papers and their associated review comments and author replies on OpenReview (publicly available via the API). For EMNLP 2024, we obtained the author response documents for papers that were eventually accepted into the main conference from NLPeer-v2~\cite{dycke-etal-2023-nlpeer}. In both cases, we also gathered the final camera-ready versions of the papers. The camera-ready is the version where we expect the promised changes to appear if fulfilled.
Table~\ref{tab:stats} summarizes basic statistics of these datasets.

\subsection{Commitment Extraction and Verification}

We define an \textbf{author commitment} as an explicit statement in the author rebuttal or response indicating that the authors commit to adding, revising, clarifying, or providing content beyond the original submission (e.g., additional experiments, analyses, clarifications, or releases).

\paragraph{Commitment Extraction}
We first employ a LLM to extract author commitments from rebuttal or response texts. Given the full author response and the associated reviewer comments, the model is prompted to identify all explicit commitments made by the authors. Importantly, the LLM extracts only the \emph{verbatim text spans} from the author response that express such commitments, without paraphrasing or rewriting.

For each extracted commitment, the model additionally identifies and aligns the corresponding reviewer comment(s) that the commitment addresses, based on topical relevance and contextual cues (e.g., referenced sections, figures, or technical terms). 
This step produces a structured list of atomic commitments paired with their related reviewer critiques.

\paragraph{Commitment Verification in the Final Version}
To assess whether a commitment is fulfilled, we again use an LLM in a verification setting. For each extracted commitment, the model is provided with:
(1) the verbatim text of a single author commitment,
(2) the final (camera-ready) version of the corresponding paper, and
(3) the associated reviewer comment as contextual information.

The model determines whether the promised change is realized in the final version and outputs a binary judgment (fulfilled or not fulfilled), optionally accompanied by a brief justification that points to concrete evidence in the final paper (e.g., sections, tables, or figures). Each commitment is evaluated independently, enabling fine-grained analysis at the level of individual author promises.

\section{Results and Analysis}
We now present our findings, covering the quantity of commitments made, the rates at which authors fulfilled them, and the characteristics of those that were not fulfilled. We compare the two conferences and examine patterns across different categories of commitments.

\subsection{Volume of Commitments: ICLR vs. EMNLP}
The first striking observation (Table~\ref{tab:stats}) is how many commitments authors make in ICLR versus EMNLP. ICLR 2025 authors made on average 11.8 commitments per paper, with a median of 11. In contrast, EMNLP 2024 authors made about 4.1 commitments on average (median 3). The maximum number of promises in a single ICLR submission was a whopping 59, whereas the highest in EMNLP was 21.

These findings have an interesting implication: reviewers and program chairs might implicitly be aware that ICLR authors will change a lot in the paper after review. Acceptance decisions may be made assuming those changes will happen. This means if those changes do \emph{not} happen, the published paper might fall short of what was promised during review. The need for auditing is thus arguably stronger in venues with long, promise-heavy rebuttals.

\begin{table}[t]
\centering
\resizebox{\columnwidth}{!}{
\begin{tabular}{lrr}
\hline
& \textbf{ICLR 2025} & \textbf{EMNLP 2024} \\
\hline
Total papers collected & 7,192 & 1,050 \\
Total commitments extracted & 84,739 & 4,320 \\
Avg. commitments per paper & 11.78 & 4.11 \\
Median commitments & 11 & 3 \\
Max commitments (single paper) & 59 & 21 \\
\hline
\end{tabular}
}
\caption{Summary statistics of author commitments identified in the rebuttals of ICLR 2025 and EMNLP 2024.}
\label{tab:stats}
\end{table}

\begin{table*}[t]
\centering
\small
    \begin{tabular}{ll|rr}
    \multicolumn{1}{c}{Severity/ Difficulty} & \multicolumn{1}{c|}{Type} & \multicolumn{1}{c}{ICLR-2025} & \multicolumn{1}{c}{EMNLP-2024} \\
    \hline
    \multirow{2}[2]{*}{High/ High} & Missing Experiments/Comparisons & 28.10\% & 31.54\% \\
          & Missing Theory/Proofs & 2.80\% & 0.94\% \\
    \hline
    \multirow{3}[2]{*}{High/ Low} & Unreleased Code/Data & 4.76\% & 2.89\% \\
          & Unclarified Concepts/Definitions & 8.10\% & 7.34\% \\
          & Missing Discussion/Limitations & 16.46\% & 17.95\% \\
    \hline
    \multirow{3}[2]{*}{Low/ Low} & Uncorrected Typos/Writing & 7.49\% & 9.29\% \\
          & Unadjusted Structure/Appendix & 8.37\% & 7.65\% \\
          & Missing Citations/References & 3.61\% & 5.85\% \\
    \hline
    Low/ High & Missing Figures/Examples & 13.56\% & 11.63\% \\
    \hline
    Unknown & Others & 6.75\% & 4.92\% \\
    \end{tabular}%
\caption{Breakdown of unfulfilled author commitments by severity, difficulty, type, and conference.}
\label{tab:broken-commitments}
\end{table*}

\subsection{Fulfillment Rates}
What fraction of the commitments were actually fulfilled in the final versions? Overall, we found that the majority of promises were kept, but a significant minority were not:
\begin{itemize}\setlength\itemsep{0em}
\item \textbf{ICLR 2025:} Approximately 75.89\% of the extracted commitments were fulfilled in the camera-ready papers, leaving about 24\% unfulfilled.
\item \textbf{EMNLP 2024:} About 72.91\% of commitments were fulfilled, with around 27\% unfulfilled.
\end{itemize}

To understand the nature of unfulfilled commitments, we classify them based on two dimensions: ``Difficulty'' (how hard the task is to execute) and ``Severity'' (how critical the failure is to the paper's scientific integrity). We propose a four-quadrant framework to categorize these broken promises, as shown in Table~\ref{tab:broken-commitments}.\footnote{The complete classification prompt and the mapping logic for categories to quadrants are detailed in Appendix~\ref{tab:promise-classification-prompt}.}

\paragraph{Quadrant I: High Severity / High Difficulty}
This category includes commitments that are central to the scientific validity of the work but are resource-intensive or technically challenging to implement.
\begin{itemize}
\small
    \item \textbf{Missing Experiments/Comparisons:} This is the most common type of broken promise across both conferences. Authors often promise additional baselines or ablations to satisfy reviewers but fail to deliver, likely due to time constraints or insufficient computational resources (e.g., running out of GPU hours).
    \item \textbf{Missing Theory/Proofs:} While less frequent, missing theoretical derivations undermines the paper's foundational claims.
\end{itemize}

\paragraph{Quadrant II: High Severity / Low Difficulty}
These commitments are critical for reproducibility and clarity but are relatively easy to execute. Failure here is particularly concerning as it may signal negligence.
\begin{itemize}
\small

    \item \textbf{Unreleased Code/Data:} Releasing code is technically straightforward but often delayed or forgotten. This raises concerns about reproducibility and potential ``vaporware.''
    \item \textbf{Unclarified Concepts/Definitions:} Simple clarifications are vital for reader understanding. Neglecting these suggests a lack of attention to detail.
    \item \textbf{Missing Discussion/Limitations:} Authors often promise to discuss limitations or broader impacts to appease reviewers but omit them in the final version, possibly to save space or avoid highlighting weaknesses.
\end{itemize}

\paragraph{Quadrant III: Low Severity / Low Difficulty}
These are minor issues that affect the paper's polish but rarely determining factors for acceptance.
\begin{itemize}
\small

    \item \textbf{Unadjusted Structure/Appendix:} A significant portion of broken promises involves reorganization, such as moving content to the appendix. Failure here often results in a cluttered manuscript.
    \item \textbf{Uncorrected Typos/Writing:} Minor errors that were promised to be fixed but remain in the final text.
\end{itemize}

\paragraph{Quadrant IV: Low Severity / High Difficulty}

These commitments improve the paper's presentation but are not strictly necessary for scientific validity.
\begin{itemize}
\small

    \item \textbf{Missing Figures/Examples:} Creating high-quality visualizations or detailed examples is time-consuming. While promised to enhance clarity, their absence is often tolerated.
\end{itemize}

This analysis highlights that while many broken promises are minor issues, a substantial number fall into the high-severity categories (Quadrants I and II), posing risks to the scientific integrity and reproducibility of the published work.

\subsection{Interpreting Non-Fulfillment and the Role of Transparency}

The results above show that a non-trivial fraction of author commitments are not fulfilled in the final, camera-ready versions. Importantly, these instances of non-fulfillment should not be interpreted simply as authors acting in bad faith or failing to honor their promises. Rather, they reflect the complex and constrained realities of the post-acceptance revision process. 

First, it is natural that commitments vary widely in feasibility and outcome. Some promised actions---such as additional experiments, ablations, or exploratory analyses---may be genuinely attempted but ultimately yield negative or inconclusive results. In such cases, non-fulfillment does not necessarily imply that the commitment was meaningless. For example, a reviewer may suggest a research direction that later turns out to be uninformative, yet the author may still have promised to explore it, and the reviewer may have trusted this promise when updating their score. Even when the outcome is negative, explicitly reporting this in the final paper (e.g., in an appendix or discussion section) would still improve transparency and better reflect the scientific process. Modern conference formats increasingly allow flexible use of appendices for such content, making it easier to accommodate these outcomes without displacing core results.

Second, the timeline between acceptance notification and camera-ready submission is often extremely short. Some commitments may be overly optimistic when made during rebuttal, especially under the pressure to respond positively to reviewer concerns. In hindsight, certain promises may turn out to be unrealistic or premature. This raises two distinct issues: on the one hand, authors should be discouraged from making commitments they cannot reasonably fulfill; on the other hand, when a promised change is not completed in time, it is reasonable to ask whether an explicit explanation should be provided in the final version, rather than silently omitting the change.

Third, some cases of non-fulfillment involve relatively minor issues, such as uncorrected typos, small wording changes, or promised structural adjustments. These are unlikely to affect acceptance decisions, but they nevertheless reduce the overall quality and polish of the published paper. A lightweight checklist that reminds authors of these commitments could help catch such oversights and improve consistency, without adding substantial burden.

Taken together, these observations suggest that the goal of auditing author commitments should not be to punish or police authors for every unfulfilled promise. Instead, the primary objective should be to improve transparency, shared understanding, and follow-through in the peer review process. By explicitly surfacing which commitments were fulfilled, which were not, and why, a structured mechanism such as an Author Commitment Checklist can support more honest rebuttals, reduce overpromising, and encourage authors to document negative results or unresolved issues when appropriate. In this sense, non-fulfillment is not merely a failure mode to be minimized, but also a signal of where clearer communication and better tooling can strengthen the integrity of peer review.

\section{Practical Feasibility of Automated Commitment Auditing}
\label{sec:feasibility}

Beyond measuring how often author commitments are fulfilled, a key question is whether an automated auditing mechanism is \emph{practically deployable} in real conference workflows. In this section, we evaluate feasibility from two complementary perspectives: (1) the reliability of LLM-based judgments relative to human reference acceptance behavior, and (2) the computational and financial cost of running such checks at scale.

\subsection{Human Reference Acceptance and Model Accuracy}

We sampled 600 commitments identified by LLMs, and these instances were annotated by native English speakers recruited via Prolific. To assess judgment consistency, we additionally included 200 overlapping instances annotated by multiple annotators.

Table~\ref{tab:agreement-accuracy} summarizes the results.
We report the \emph{reference acceptance rate}, defined as the proportion of annotator-provided reference answers that are judged as correct by an evaluator.
Human evaluators achieve a reference acceptance rate of 72.86\%, and LLM-based evaluators exhibit comparable acceptance behavior.
Across three models, acceptance rates range from 74\% to 78\%, with relatively small differences between models.

\begin{table}[t]
\centering
\small
\begin{tabular}{lr}
\hline
\textbf{Metric} & \textbf{Score} \\
\hline
Human Reference Acceptance Rate & 72.86\% \\
\hline
Claude-Sonnet-4.5 & 75.38\% \\
Gemini-2.5-Pro & 78.00\% \\
GPT-5 & 74.05\% \\
\hline
\end{tabular}
\caption{Human reference acceptance rate and LLM evaluation accuracy.}
\label{tab:agreement-accuracy}
\end{table}

Two observations are worth highlighting. First, the relatively modest reference acceptance rate among humans indicates that perfect accuracy should not be expected, even from an ideal system. Second, achieving approximately 75\% accuracy already ensures that most authors would not be burdened by spurious or excessive false alarms. As models continue to improve and as more annotated data becomes available for calibration, we expect these results to further improve.

Importantly, these findings also motivate a design choice for deployment: the auditing output should explicitly allow for uncertainty. In particular, an Author Commitment Checklist should include options such as \emph{``not applicable''}, \emph{``commitment no longer holds''}, or \emph{``possible detection error''}, allowing authors to contest or clarify ambiguous cases rather than treating the model output as a hard enforcement mechanism.

\subsection{Computational and Financial Cost}

A second concern is whether automated commitment auditing is affordable at the scale of modern conferences. We estimate the cost by measuring the average number of input and output tokens required to (1) identify commitments and (2) verify them against the camera-ready paper.

We first consider paper length. Table~\ref{tab:paper-length} shows that EMNLP 2024 papers have an average length of 17.54 pages, with a maximum of 62 pages, while ICLR papers are substantially longer. These lengths determine the token budget required for LLM-based analysis.

\begin{table}[t]
\centering
\small
\begin{tabular}{lrr}
\hline
 & \textbf{ICLR 2025} & \textbf{EMNLP 2024} \\
\hline
Average length (pages) & 27.89 & 17.54 \\
Maximum length (pages) & 150 & 62 \\
\hline
\end{tabular}
\caption{Paper length statistics for the two conferences.}
\label{tab:paper-length}
\end{table}

Focusing on EMNLP 2024 as a representative case, Table~\ref{tab:cost} reports the average token usage and estimated cost per paper for the two main steps in our pipeline.
Specifically, we use \texttt{gemini-2.5-Pro} for promise identification (input: \$1.25 / 1M tokens; output: \$10 / 1M tokens), and \texttt{gemini-3-preview} for commitment verification (input: \$2 / 1M tokens; output: \$12 / 1M tokens).
Even when accounting for both stages, the total cost per paper is approximately \$0.0278 USD.

From a practical standpoint, this cost is negligible relative to typical conference budgets. For example, it could be absorbed by a marginal increase in the paper registration fee, or simply treated as operational overhead. Even at large-scale venues with thousands of accepted papers, the total cost would remain modest and would not materially affect conference finances.

\subsection{Implications for Deployment}

Taken together, our findings suggest that automated auditing of author commitments is both technically and economically feasible. While LLM-based judgments are imperfect, their accuracy approaches human agreement, and their errors can be mitigated through careful interface design that emphasizes transparency rather than enforcement. Meanwhile, the computational cost is sufficiently low to allow routine deployment without imposing meaningful financial burden.

These results support the viability of integrating an Author Commitment Checklist into existing camera-ready workflows as a lightweight yet impactful intervention. Rather than introducing new reviewer labor or strict policing, such a system would primarily serve as a reminder and documentation tool, nudging authors toward follow-through and clearer communication. In doing so, it offers a pragmatic path toward improving accountability in peer review without undermining its collaborative spirit.

\begin{table}[t]
\centering
\resizebox{\columnwidth}{!}{
\begin{tabular}{lcc}
\hline
 & \textbf{Promise Identification} & \textbf{Commitment Verification} \\
\hline
Avg. input tokens & 2,875.82 & 5,384.24 \\
Avg. output tokens & 713.56 & 525.27 \\
\hline
Input cost (USD) & 0.0036 & 0.0108 \\
Output cost (USD) & 0.0071 & 0.0063 \\
\hline
\multicolumn{3}{c}{\textbf{Total cost per paper: \$0.0278 USD}} \\
\hline
\end{tabular}
}
\caption{Estimated LLM token usage and cost per paper (EMNLP 2024).}
\label{tab:cost}
\end{table}

\begin{table*}[t]
\centering
\footnotesize
\renewcommand{\arraystretch}{1.15}
\begin{tabular}{p{0.58\linewidth} p{0.36\linewidth}}
\hline
\textbf{Checklist Item} & \textbf{Author Confirmation} \\
\hline
\textbf{``We will evaluate on MovieCoref and discuss other suggested datasets in the final version.''} \newline
{\em Type:} Missing experiments/comparisons (high severity) \newline
{\em System check:} not fulfilled \newline
{\em Evidence:} ``MovieCoref'' not found; experiments only on LitBank/FantasyCoref/AFT.
&
\checkbox\ Fulfilled (point to section/table) \newline
\checkbox\ Not fulfilled (briefly explain why) \newline
{\em Reason:} resource constraints / out of scope / space / other \newline
\checkbox\ System error (model misdetection) \\
\hline
\textbf{``We will make the hallucination sentence more concise.''} \newline
{\em Type:} Writing revision (low severity) \newline
{\em System check:} no change detected \newline
{\em Evidence:} Sentence unchanged between rebuttal and final version.
&
\checkbox\ Fulfilled \newline
\checkbox\ Not fulfilled (optional note) \newline
\checkbox\ System error (model misdetection) \\
\hline
\end{tabular}
\caption{Illustrative author-facing Author Commitment Checklist (ACC). Authors explicitly confirm fulfillment, explain non-fulfillment, or report system misdetection, reflecting the non-binding and fallible nature of automated checks.}
\label{tab:acc-ui}
\end{table*}

\section{Author Commitment Checklist}

Our results suggest that non-fulfillment is often driven by (i) \textbf{unforced errors} (high-severity but low-difficulty omissions such as missing code release or missing clarifications) and (ii) \textbf{structural neglect} (low-severity housekeeping items such as promised rewording or restructuring) rather than deliberate misconduct. In Section \ref{sec:feasibility}, we further show that LLM-based verification is sufficiently accurate and affordable to be deployed as a lightweight reminder-and-documentation mechanism. Building on these insights, we propose an \textbf{Author Commitment Checklist (ACC)}: a structured checklist that records commitments made during rebuttal and prompts authors to either (a) provide evidence that the promise is fulfilled in the camera-ready version, or (b) provide a brief, explicit justification for why it is not.

\subsection{Design Principles: Transparency over Punishment}
ACC is not intended to ``police'' authors or introduce additional reviewer labor. Instead, it operationalizes the transparency goal discussed in \S\ref{sec:feasibility}: if a commitment is not implemented (e.g., because the experiment was negative, infeasible, or deprioritized), the final record should \emph{say so explicitly} rather than silently omitting it. Concretely, ACC is designed around three principles:

\begin{itemize}\setlength\itemsep{0em}
    \item \textbf{Make commitments observable:} surface the exact commitment text (verbatim) and the reviewer request it addresses.
    \item \textbf{Make fulfillment verifiable:} require either a pointer to where it appears in the final paper (section/table/figure) or a short justification for non-fulfillment.
    \item \textbf{Prioritize high-impact items:} emphasize Quadrant I/II items (high severity) to reduce the risk of publishing papers that fall short of what reviewers expected at decision time.
\end{itemize}

\subsection{Workflow Integration}

ACC is designed to integrate into existing camera-ready workflows (e.g., OpenReview, Softconf) with minimal friction. Importantly, the workflow is \emph{trust-first by default}: authors are primarily responsible for verifying and reporting the status of their own commitments. At the same time, the design remains \emph{audit-ready}, allowing conferences to enable additional verification steps when stronger accountability is desired.

\paragraph{Step 1: Commitment extraction (pre-decision).}
Prior to decision notification, an automated system extracts atomic author commitments verbatim from the author responses and optionally aligns them with the corresponding reviewer comments. The extracted list can be shared with authors as a reference, allowing them to review the commitments they have articulated before acceptance decisions are finalized.

\paragraph{Step 2: Author self-verification (pre-camera-ready).}
Before the camera-ready deadline, authors run the ACC verification tool on their own draft camera-ready paper. The tool produces a tentative fulfillment report for each extracted commitment, together with supporting evidence (or an indication that no evidence was detected). Authors review this report and submit it alongside the camera-ready version as a self-check. This step is lightweight and non-binding, and is intended to reduce accidental non-fulfillment rather than to enforce compliance.

\paragraph{Step 3: Optional official verification (post-camera-ready).}
For venues that require stronger auditing guarantees, the conference system may optionally run the same ACC verification procedure on the submitted final papers after the camera-ready deadline. The resulting checklist can be returned to authors for confirmation, allowing them to (i) confirm fulfillment, (ii) briefly explain non-fulfillment, or (iii) report a system misdetection with a pointer to the relevant section. This step is not required for all venues and is primarily intended to support auditing and consistency checks rather than to override author judgment.

\paragraph{Step 4: Organizer-facing summary (optional).}
Program chairs may optionally view aggregate summaries (e.g., counts of unfulfilled high-severity commitments) to support shepherding, conditional acceptance, or post-hoc analysis, depending on the conference’s governance model.

\subsection{What Authors Actually See: Concrete Examples}

For ACC to be effective, it is crucial that commitments are presented to authors in a form that is \emph{actionable}, \emph{interpretable}, and aligned with existing camera-ready workflows. Rather than exposing raw model outputs or machine-readable representations, ACC is deliberately designed as a human-facing checklist. Each checklist item makes a single commitment explicit, summarizes the system’s tentative assessment, and requires the author to actively confirm fulfillment or provide a brief explanation for non-fulfillment.

Table~\ref{tab:acc-ui} illustrates how ACC would appear to authors at camera-ready submission time. Each row corresponds to one atomic commitment extracted verbatim from the rebuttal. The system-provided status is intended as a reminder and diagnostic signal rather than a binding judgment; authors can override it, but must do so explicitly.

This form-based presentation serves two purposes. First, it reduces accidental non-fulfillment caused by oversight or time pressure, particularly for low-difficulty items such as promised clarifications or writing edits. Second, when a commitment cannot be fulfilled (e.g., due to negative experimental results or resource constraints), it encourages authors to document this explicitly rather than omitting the change without explanation. Importantly, ACC does not aim to penalize authors for non-fulfillment; instead, it makes the outcome of rebuttal commitments transparent and auditable, strengthening the alignment between review-time discussion and the final published record.

\subsection{Anticipated Impact and Scope}
ACC primarily targets Quadrant II and Quadrant III failures, where omissions are both common and preventable via reminders and structured documentation. By requiring either evidence or a brief justification, ACC can reduce silent non-fulfillment, improve the fidelity between rebuttal discourse and published record, and discourage overpromising during rebuttal. Importantly, ACC remains compatible with diverse conference norms: venues can start with a purely author-facing reminder (no enforcement) and later adopt optional organizer-facing summaries for shepherding or conditional acceptance if desired.

\section{Discussion and Conclusion}
\label{sec:conclusion}

This paper argues for a simple but missing piece in current peer review workflows: making rebuttal-time promises \emph{visible, checkable, and interpretable} at camera-ready time. Our empirical audit shows that author commitments are common and that a non-trivial portion remain unfulfilled, including high-severity items such as missing experiments or unreleased artifacts. At the same time, we emphasize that non-fulfillment is not automatically evidence of bad faith: negative results, feasibility constraints, and short camera-ready timelines all make certain commitments difficult to complete. This is precisely why we frame ACC as a transparency and documentation mechanism rather than a punitive enforcement tool.

Importantly, ACC is designed to be lightweight in the typical case. For papers with roughly 4--11 commitments on average (Table~\ref{tab:stats}), confirming a short checklist is a small marginal burden compared to writing a rebuttal and preparing the final revision. The larger burden in the current system often comes from a different source: reviewers frequently request many additional experiments, and prior work suggests that such requests can be driven by limited time and heuristic-driven reviewing behavior~\cite{purkayastha-etal-2025-lazyreview}. Under these conditions, it is easy for the author--reviewer interaction to drift toward a pattern where authors feel compelled to promise ``more'' in order to secure a better score, even when the requested additions are only weakly connected to the paper's core claims.

Seen from this perspective, ACC is not only about accountability; it is also a tool for reflection on what rebuttal commitments should mean. By requiring authors to either point to evidence of fulfillment or explicitly explain non-fulfillment, ACC encourages the community to distinguish between (i) requests that genuinely affect the paper's claims and (ii) requests that mainly inflate workload without changing the scientific conclusion. This, in turn, can support healthier norms: authors can respond more confidently and precisely to reviewer concerns when the discussion is anchored in claims and evidence, rather than in score-driven negotiation.

More broadly, we should be cautious about optimizing the peer review process for numerical ratings at the expense of communication quality. We are doing research, not ``scorewashing.'' Designing interaction mechanisms that make rebuttal and revision more meaningful is an important community problem. ACC is a small, practical step in this direction: it nudges the system toward clearer commitments, more honest follow-through, and better alignment between review-time discourse and the published record.

\section*{Limitation}
Our study has several limitations that should be considered when interpreting the results and the proposed Author Commitment Checklist (ACC).

\paragraph{Coverage and representativeness.}
We analyze commitments from only two venues (ICLR 2025 and EMNLP 2024). These conferences differ in transparency, the time length of the rebuttal, and response format, but they may not represent other fields, venues with different rebuttal policies (e.g., no rebuttal, shepherding-only), or journals. 

\paragraph{Definition and granularity of “commitments.”}
We operationalize commitments as \emph{explicit} action statements in rebuttals. This choice misses implicit intentions (e.g., “we agree this is important” without a clear action), and it can also split or merge actions in imperfect ways (e.g., one sentence containing multiple sub-promises). As a result, the reported counts and fulfillment rates depend on the extraction granularity and the boundary between “promise” vs.\ “discussion.”

\paragraph{LLM extraction and verification errors.}
Both commitment extraction and fulfillment verification rely on LLM judgments, which are fallible. Errors include: (i) false positives where the model flags non-promissory text as a commitment; (ii) false negatives where subtle promises are missed; (iii) misalignment of a promise to the wrong reviewer comment; and (iv) verification mistakes when the evidence is paraphrased, relocated to an appendix, presented in a figure, or expressed indirectly. The human agreement rate (Table~\ref{tab:agreement-accuracy}) indicates that the task itself is ambiguous even for humans, so model error is not merely an implementation issue but partly reflects intrinsic subjectivity.

\paragraph{Fulfillment as a binary label.}
We treat fulfillment as a binary outcome, but many real cases are partial: authors may run some but not all promised experiments, add a discussion without the depth implied, or provide code that is incomplete or non-functional. Binary scoring can therefore overstate both compliance (counting minimal changes as fulfilled) and non-compliance (counting partial progress as not fulfilled). A more faithful evaluation would use graded labels (fulfilled / partially / not fulfilled) and separate “attempted but negative result” from “not attempted.”

\paragraph{Interpretation and normative scope.}
Non-fulfillment should not be equated with bad faith. Some promises become infeasible, out of scope, or yield negative results. Our analysis does not estimate intent, effort, or the decision-criticality of each promise for acceptance. Accordingly, ACC should be viewed as a transparency and documentation mechanism rather than an enforcement tool, and deployment should be careful to avoid incentivizing superficial “box-checking” edits.

\paragraph{Potential behavioral effects of ACC.}
While ACC could reduce accidental omissions, it may also change rebuttal behavior: authors might write fewer explicit commitments, hedge language to avoid being audited, or shift promises into vague statements. Evaluating these second-order effects requires a prospective study (e.g., an A/B test at a venue) that we do not conduct here.

\section*{Acknowledgment}
This paper is based on results obtained from AIST policy-based budget project "R\&D on Generative AI Foundation Models for the Physical Domain".
This work has been co-funded by the European Union (ERC, InterText,
101054961) and the LOEWE Distinguished Chair “Ubiquitous Knowledge Processing”, LOEWE initiative, Hesse, Germany (Grant
Number: LOEWE/4a//519/05/00.002(0002)/81). Views and opinions expressed are
however those of the author(s) only and do
not necessarily reflect those of the European
Union or the European Research Council. Neither the European Union nor the granting authority can be held responsible for them.

\bibliography{custom}

\appendix
\section{Prompt}
Table~\ref{tab:promise-extraction-prompt} provides the prompt that we used to identify the author commitments, and Table~\ref{tab:promise-verification-prompt} shows the prompt used to verify the commitment. 

\begin{table*}[t]
\centering
\small
\renewcommand{\arraystretch}{1.2}
\begin{tabular}{p{0.25\linewidth} p{0.7\linewidth}}
\hline
\textbf{Section} & \textbf{Prompt Content} \\
\hline
Task Description &
You are given two inputs from an OpenReview discussion: \newline
(1) AUTHOR\_RESPONSE: \newline
\texttt{\{author\_response\}} \newline
(2) REVIEWER\_COMMENTS (each bracketed and numbered): \newline
\texttt{\{joined\_comments\}} \\
\hline
Task &
Extract author \textbf{promises} and the \textbf{reviewer comments} they correspond to. \\
\hline
Definition of a Promise &
A \emph{promise} is any explicit author commitment, clarification, or action statement — for example: \newline
• “We have added …” \newline
• “We will add / revise / clarify / release …” \newline
• “We have reworded …” \newline
• “We now include …” \newline
• “We push this to future work …” \newline
• “We added references / figures / experiments …” \newline
Extract only the \textbf{exact verbatim text spans} from the AUTHOR\_RESPONSE that express these promises. \\
\hline
Definition of an Aligned Reviewer Comment &
For each promise, extract one or more \textbf{reviewer comment snippets} (also verbatim, from REVIEWER\_COMMENTS) that the promise addresses or responds to. \newline
• Each reviewer snippet should include enough context to understand the critique or request. \newline
• If multiple reviewer comments are relevant, include all. \newline
• If none clearly align, use an empty list for that promise. \\
\hline
Matching Guidance &
• Match by topic or cue words (e.g., section numbers, figure mentions, algorithm names, terms like “DAS”, “OOD”, “Same-Object”, etc.). \newline
• Prefer concise but complete spans (paragraph level). \newline
• Do not paraphrase or rewrite anything. \newline
• Preserve original punctuation and line breaks. \\
\hline
Strict Output Format &
Return only valid JSON with the exact schema below, and nothing else: \newline
\texttt{\{} \newline
\texttt{  "promises": [ "..."],} \newline
\texttt{  "aligned\_reviewer\_comments": [[ "..."]]} \newline
\texttt{\}} \\
\hline
Constraints &
• The number of elements in \texttt{"promises"} and \texttt{"aligned\_reviewer\_comments"} must match. \newline
• Each element in \texttt{"aligned\_reviewer\_comments"} is a list (possibly empty). \newline
• Output only the JSON block, no explanations, no markdown. \\
\hline
\end{tabular}
\caption{The prompt used to extract author commitment and their aligned reviewer comments.}
\label{tab:promise-extraction-prompt}
\end{table*}

\begin{table*}[t]
\centering
\small
\renewcommand{\arraystretch}{1.2}
\begin{tabular}{p{0.25\linewidth} p{0.7\linewidth}}
\hline
\textbf{Section} & \textbf{Prompt Content} \\
\hline
Role Definition &
You are an expert academic reviewer. \\
\hline
Inputs &
You are provided with: \newline
• A PDF file of a Final Paper \newline
• A list of Author Promises made during the rebuttal \newline
• Reviewer comments as contextual information to understand the intent of each promise \\
\hline
Task &
Check if each promise was fulfilled in the provided PDF file. \\
\hline
Promise List &
PROMISE LIST: \newline
\texttt{\{json.dumps(items, indent=2, ensure\_ascii=False)\}} \\
\hline
Evaluation Objective &
Determine whether the content claimed in each author promise actually appears in the final paper PDF, based on explicit evidence (e.g., sections, tables, figures, experiments, or textual changes). \\
\hline
Output Format &
Return ONLY a valid JSON Array. \newline
Example: \newline
\texttt{[} \newline
\texttt{  \{} \newline
\texttt{    "promise\_text": "We will add an ablation study...",} \newline
\texttt{    "is\_fulfilled": true,} \newline
\texttt{    "fulfillment\_evidence": "Table 4 in Section 5 shows..."} \newline
\texttt{  \}} \newline
\texttt{]} \\
\hline
Output Constraints &
• Output must be a valid JSON array \newline
• Each element corresponds to one promise \newline
• No explanations, markdown, or extra text outside the JSON output \\
\hline
\end{tabular}
\caption{The prompt used to verify whether the author commitment made during rebuttal are fulfilled in the final paper.}
\label{tab:promise-verification-prompt}
\end{table*}

\section{Prompts for Commitment Classification}
Table~\ref{tab:promise-classification-prompt} provides the prompt used to classify unfulfilled promises into specific categories and determine their severity and difficulty quadrants.

\begin{table*}[t]
\centering
\small
\renewcommand{\arraystretch}{1.2}
\begin{tabular}{p{0.25\linewidth} p{0.7\linewidth}}
\hline
\textbf{Section} & \textbf{Prompt Content} \\
\hline
Role Definition &
You are an expert academic reviewer and meta-reviewer. \\
\hline
Task Description &
Classify the following author commitment (from an academic paper revision plan) into one or more categories based on the promise text and the evidence of fulfillment failure. \\
\hline
Categories &
Choose ONLY from this list: \newline
• Missing Experiments/Comparisons \newline
• Unreleased Code/Data \newline
• Uncorrected Typos/Writing \newline
• Missing Citations/References \newline
• Missing Figures/Examples \newline
• Missing Theory/Proofs \newline
• Missing Discussion/Limitations \newline
• Unadjusted Structure/Appendix \newline
• Unclarified Concepts/Definitions \newline
• Others \newline
• Not a Promise \\
\hline
Inputs &
\textbf{Promise Text:} \newline
\texttt{"\{promise\_text\}"} \newline
\textbf{Fulfillment Evidence (Why it failed):} \newline
\texttt{"\{evidence\}"} \\
\hline
Rules &
1. Multi-label allowed if the promise covers multiple distinct actions. \newline
2. If the text is a question to the reviewer or lacks a concrete commitment, use "Not a Promise". \newline
3. Return STRICT JSON format. \\
\hline
Output Format &
Return only valid JSON with the following schema: \newline
\texttt{\{} \newline
\texttt{  "is\_promise": true,} \newline
\texttt{  "categories": ["Category Name"],} \newline
\texttt{  "rationale": "one sentence explanation"} \newline
\texttt{\}} \\
\hline
\end{tabular}
\caption{The prompt used to classify unfulfilled author commitments into the nine defined categories.}
\label{tab:promise-classification-prompt}
\end{table*}

\section{Quadrant Mapping Logic}
The classification results from Table~\ref{tab:promise-classification-prompt} are mapped to a four-quadrant framework to assess the impact of broken promises. The mapping logic used in our analysis is detailed in Table~\ref{tab:quadrant-mapping}.

\begin{table*}[h]
\centering
\small
\renewcommand{\arraystretch}{1.2}
\begin{tabular}{l|l|l}
\hline
\textbf{Category} & \textbf{Quadrant} & \textbf{Description} \\
\hline
Missing Experiments/Comparisons & Q1 & High Severity / High Difficulty \\
Missing Theory/Proofs & Q1 & High Severity / High Difficulty \\
\hline
Unreleased Code/Data & Q2 & High Severity / Low Difficulty \\
Unclarified Concepts/Definitions & Q2 & High Severity / Low Difficulty \\
Missing Discussion/Limitations & Q2 & High Severity / Low Difficulty \\
\hline
Uncorrected Typos/Writing & Q3 & Low Severity / Low Difficulty \\
Missing Citations/References & Q3 & Low Severity / Low Difficulty \\
Unadjusted Structure/Appendix & Q3 & Low Severity / Low Difficulty \\
\hline
Missing Figures/Examples & Q4 & Low Severity / High Difficulty \\
\hline
\end{tabular}
\caption{Mapping logic from categories to the four-quadrant framework (Severity vs. Difficulty).}
\label{tab:quadrant-mapping}
\end{table*}

\end{document}